\documentclass[structabstract]{aa}
\usepackage{natbib}
\usepackage{graphicx}
\usepackage{subfigure}

\begin{document}

\title{Thermal structure of hot non-flaring corona from Hinode/EIS}

\author{A. Petralia\inst{1} \and F. Reale\inst{1}\inst{2} \and P. Testa\inst{3} \and G. Del Zanna\inst{4}}

\institute{Dipartimento di Fisica \& Chimica, Universit\`a di Palermo, Piazza del Parlamento 1, 90134, Italy\\
\and INAF-Osservatorio Astronomico di Palermo, Piazza del Parlamento 1, 90134 Palermo, Italy\\
\and Harvard-Smithsonian Center for Astrophysics, 60 Garden Street, MS 58, Cambridge, MA 02138, USA \\
\and DAMTP, Centre for Mathematical Sciences, Wilberforce Road, Cambridge, UK}

\abstract{}
{In previous studies a very hot plasma component has been diagnosed in solar active regions through the images in three different narrow-band channels of SDO/AIA. This diagnostic from EUV imaging data has also been supported by the matching morphology of the emission in the hot Ca XVII line, as observed with Hinode/EIS. This evidence is debated because of unknown distribution of the emission measure along the line of sight. Here we investigate in detail the thermal distribution of one of such regions using EUV spectroscopic data.}
{In an active region observed with SDO/AIA, Hinode/EIS and XRT, we select a subregion with a very hot plasma component and another cooler one for comparison. The average spectrum is extracted for both, and 14 intense lines are selected for analysis, that probe the $5.5 < \log T < 7$ temperature range uniformly. From these lines the emission measure distributions are reconstructed with the MCMC method. Results are cross-checked with comparison of the two subregions, with a different inversion  method, with the morphology of the images, and with the addition of fluxes measured with from narrow and broad-band imagers.}
{We find that, whereas the cool region has a flat and featureless distribution that drops at temperature $\log T \geq 6.3$, the distribution of the hot region shows a well-defined peak at $\log T = 6.6$ and gradually decreasing trends on both sides, thus supporting the very hot nature of the hot component diagnosed with imagers. The other cross-checks are consistent with this result.}
{This study provides a completion of the analysis of active region components, and the resulting scenario supports the presence of a minor very hot plasma component in the core, with temperatures $\log T > 6.6$. }


\authorrunning{Petralia et al.}
\titlerunning{Hinode/EIS hot non-flaring corona}

\maketitle

\section{Introduction}
It is accepted that the energy source that sustains the high temperature of solar corona is in the magnetic field \citep{2006SoPh..234...41K}. Regarding the way in which this energy is converted to thermal energy, we can broadly distinguish between two scenarios that depend on the timescale of the energy release: one is continuous heating, the other is in the form of discrete, rapid pulses. The latter is consistent with the nanoflare model proposed by  \cite{1988ApJ...330..474P}. According to this model, the magnetic field tubes are displaced by the photospheric motions, and can approach and interact. When two flux tubes are almost in contact they will form a current sheet, where the field lines can reconnect. The reconnection can release a large quantity of energy in impulsive events called nanoflares. Signatures of these events are difficult to observe for several reasons, one of which is the fine structuring of the magnetic loops, that is hardly resolved with present-day instruments (e.g., \citealt{2013ApJ...770L...1T}). 

One feature to discriminate the heating release is the presence or absence of very hot plasma. In active regions, the mean coronal temperature is typically 2-3 MK. If heat pulses occur, we expect that a small amount of plasma hotter than the average (6-10 MK) will be ever-present. 

Recently, a number of studies, mostly based on data from the Hinode and the Solar Dynamics Observatory (SDO) missions, have shown increasing evidence for such small very hot components in active regions (\citealt{2009ApJ...704L..58R}, \citealt{2009ApJ...698..756R}, \citealt{2009ApJ...697...94M}, \citealt{2009ApJ...693L.131S}, \citealt{2009ApJ...697.1956K}, \citealt{2009ApJ...696..760P}, \citealt{2010AstL...36...44S}, \citealt{2010A&A...514A..82S}), but the issue is still debated (\citealt{2012ApJ...754L..40T}, \citealt{2011ApJ...734...90W}). In SDO images taken with a channel sensitive also to emission of plasma at 6 MK (94 \AA), cores of active regions contain bright strands \citep{2011ApJ...736L..16R}, as predicted by models of nanoflaring loops \citep{2010ApJ...719..576G}. It remains to be proven whether this plasma is really at such high temperatures or not. Spectroscopic observations should help greatly. In this work, we analyse an active region that shows evidence for this very hot component in SDO data, but for which spectroscopic data are also available from the EUV spectrometer EIS on-board the Hinode mission. In a previous work  (\citealt{2012ApJ...750L..10T}, hereafter Paper I), emission in the CaXVII line, which forms around temperature of 6-8MK, was detected in the hot structures identified with SDO/AIA data. In that work they built a 3 color image, to highlight the presence of a very hot component of emitting plasma inside the active region. The AIA 94~\AA\ band  is known to be multi-thermal.
 It is sensitive to hot plasma, due to the presence of an 
Fe XVIII line, formed around 6~MK, but is also sensitive to plasma at 1 MK, because of the presence of 
a Fe X line and  cooler Fe IX and Fe VIII (see, e.g., \citealt{2011A&A...535A..46D}, \citealt{2012ApJ...745..111T}, \citealt{2011ApJ...743...23M}, \citealt{2011ApJ...740L..52F}, \citealt{2012A&A...537A..22O}, \citealt{2012A&A...546A..97D}).
Recently, also a Fe XIV line was identified in  \cite{2012A&A...546A..97D}. This line is normally stronger than the other cool components in 
active region cores, as shown in \cite{2013A&A...558A..73D}. It is therefore not simple to assess if the 
hot emission seen in the AIA 94~\AA\ band is really due to Fe XVIII. To clarify this point, \cite{2012ApJ...750L..10T}
 compared the AIA 94~\AA\ image with the Ca XVII image obtained from the EIS spectrometer. They showed a strong correlation between the hot CaXVII and the emission in the 94\AA\ AIA band, so concluded that the hot emission seen in the  AIA 94~\AA\ band 
is effectively due to very hot plasma (6-8MK). More direct evidence has been found from observations of another Fe XVIII line by the  SUMER spectrometer on board the SOHO mission \citep{2012ApJ...754L..40T}.

However, even if the indication is rather strong, it is still not enough to establish that the plasma is actually so hot, since in theory it is possible that plasma at a lower temperature, but with very high emission measure, can give the same line intensity, as suggested by 
\cite{2012ApJ...754L..40T}. Indeed  \cite{2013A&A...558A..73D} used simultaneous EIS and AIA observations of 
active regions cores to show that a significant fraction of the Fe XVIII 94~\AA\ intensity can be due to plasma  at 3~MK and not
6~MK. In fact,  \cite{2013A&A...558A..73D} showed that often Fe XVIII 94~\AA\ emission is present in the cores of active regions, 
but Ca XVII (which has a narrower formation temperature, hence is sensitive to hotter plasma) is not.
The only way to disentangle the various contributions to the AIA 94~\AA\ band  is therefore  to perform an emission measure modelling.
To this purpose, here we use the same observations from Hinode and SDO as in Paper I, that include both high-resolution spectroscopic data, over a wide spectral window, and images with high spatial resolution. 

All this information provides simultaneous constraints on the plasma thermal structure along the line of sight. To further support the analysis we replicate the same analysis  on the same data set but taking a region outside of the core that shows no evidence of these very hot components.We also compare two different inversion methods.
In Section~\ref{sec:obs} we describe the observation and the data analysis, and in Section~\ref{sec:discuss} the results are discussed.

\section{Observation and data analysis}
\label{sec:obs}

We analyse the active region (AR 11289) observed on 2011-09-13, from 10.30 to 11.30 UT. Our analysis is focused on spectroscopic data of EIS on board of Hinode \citep{2007SoPh..243...19C}. We analyse data from a study designed by one of us (GDZ), called ATLAS\_60, where the entire full spectral range, 178-213 \AA\ and 245-290 \AA, is extracted. 
The observations are obtained by stepping the 2$"$ slit from solar west to east, with a 120$"$x160$"$ field of view. The exposure time was 60 s. The EIS data were processed using eis\_prep, available in SolarSoft. This routine removes CCD dark current, cosmic ray strikes and takes into account hot, warm and dusty pixels. After that, radiometric calibration was applied to convert digital data (related to photon counts) into physical units (erg s$^{-1}$ cm$^{-2}$ sr$^{-1}$ \AA). 
We then re-aligned the fields of view by correcting for the wavelength offset of the two CCDs (by using the EIS routine eis\_ccd\_offset), obtained a new cropped field of view 120$" \times 140"$.

\begin{figure*}[t!]
	\centering
	\includegraphics[scale=0.8]{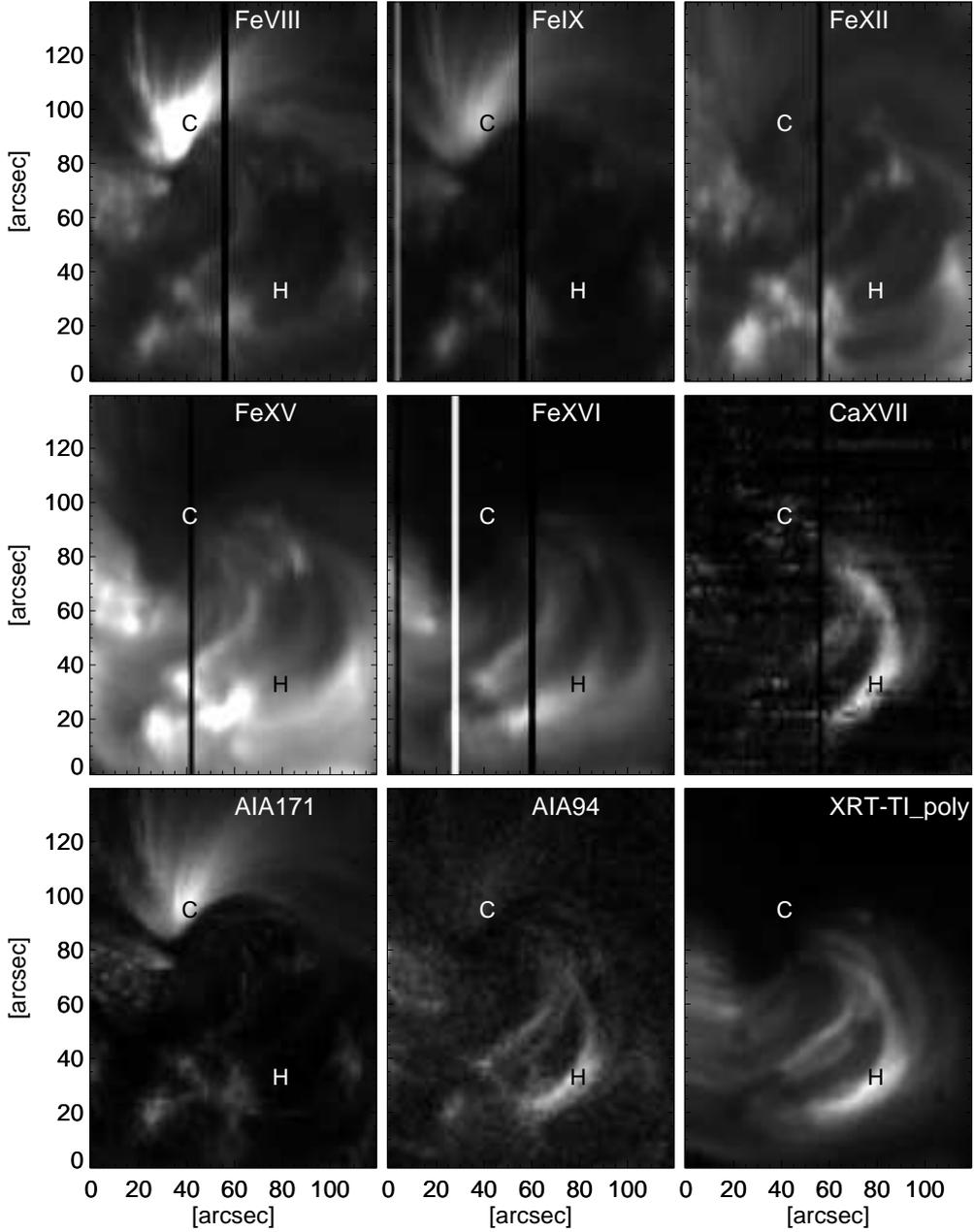}  
	\caption{EIS, AIA and XRT images of the analysed region, i.e. 6 EIS spectral lines (Fe VIII, Fe IX, Fe XII, Fe XV, Fe XVI, Ca XVII, see Table~\ref{tab:eislines}), 2 AIA channels (171 \AA, 94 \AA), and 1 XRT channel (Ti\_poly). The position of the hot (H) and cold (C) regions analysed here are marked in all images.}
	\label{fig:lines}
\end{figure*}

In the same temporal window we selected frames from two imagers, the X-ray telescope (XRT, \citealt{2007SoPh..243...63G}) on board Hinode (512$"$x512$"$ in Ti\_poly filter, exposure times between 0.7 and 1 s) and the Atmospheric Imaging Assembly (AIA, \citealt{2012SoPh..275...17L}) on board SDO (full disk in 171 \AA, 335 \AA, 94 \AA\ channels, exposure times 2, 2.9, 2.9 s, respectively) . 

The two sets of images were processed with the standard routines available in the Solar Soft package (xrt\_prep and aia\_prep). The images of the two instruments were co-aligned to each other (tr\_get\_disp.pro in SolarSoftware package) and to the EIS raster image in the He II 256 \AA\ line, to match the EIS field of view. To improve the homogeneity between the XRT or AIA images and the EIS images we have to consider that the latter are built from rastering that takes some time, while the former are "instantaneous". We then built-up new composite XRT and AIA images, that account for this different time spacing, as follows: each vertical strip is extracted from the images closest in time to the time when the EIS slit was at that location. XRT errors were computed taking into account the effect of the deterioration of the CCD response \citep{2013arXiv1312.4850K}. Fig.\ref{fig:lines} shows EIS, AIA and XRT representative images of the same field of view and with the same timing. The EIS images are obtained by integrating the spectrum in a narrow band (0.1 \AA wide) centered at the position of each line.

The information available for very hot plasma from EIS data is mostly based on the Ca XVII line, the strongest EIS line formed around 6 MK \citep{2008A&A...481L..69D}. This line is severely blended with Fe XI and O V lines, as discussed in \cite{2007PASJ...59S.857Y}, \cite{2009A&A...508..501D}, \cite{2011A&A...526A...1D}, \cite{2012ApJ...750L..10T}. As in Paper I in the case of the Ca XVII line we use the procedure developed by \cite{2009ApJ...697.1956K} for de-blending the line from Fe XI and O V lines, and extract the Ca XVII emission.

AIA and XRT images are composite images as explained above. The line images span emission from plasma in a broad temperature range $5.7 < \log T < 6.7$. The "cold" ones ($\log T < 6$) show a very bright feature consisting of "fan" loops in the top-left region, consistently with the image in the AIA 171 \AA\ channel. In the images around $\log T = 6.4$ the morphology becomes very different: the whole region in the bottom part is quite bright and closed loops are visible, while where the bright "cold" features are present there is little emission.
That hot and cold emission is often not co-spatial has been known for a long time, and is often true even at the 
high spatial resolution of AIA \citep{2013A&A...558A..73D}.
The Ca XVII image shows only two bright loops bifurcating southwards from a common point near the middle of the field of view. 
These features clearly show similar morphology to the AIA 94 \AA\ emission, and are the same features found in Paper I. Here we also analyze XRT observations, which show that these loops are also bright in the X-rays.
We see also some other emission around that recalls the EIS Fe XV and Fe XVI images.

\begin{center}
\begin{table*}[htb!]
\begin{center}
\caption{EIS line fluxes measured after fitting each selected line with a Gaussian profile, in the two regions (cold, hot) marked in Fig.\ref{fig:aia3col}. The ratio of the model to the observed fluxes are also reported for the DEM reconstructions with the MCMC and Del Zanna (DZ) methods (see Section~\ref{subsec:DEMrec}).}
\begin{tabular}{lccccccccc} 
\hline\hline
 {\bf Line}	&   Wavelength 	&	Temperature & Cold region &Ratio &Ratio &Hot region &Ratio &Ratio \\
  	&   (\AA)	&	 ($\log (T)$)& (erg cm$^{-2}$ sr$^{-1}$ s$^{-1}$)&DZ&MCMC&(erg cm$^{-2}$ sr$^{-1}$ s$^{-1}$)&DZ&MCMC\\
  \hline
	FeVIII & $185.21$ & $5.7$ & $4250 \pm 80$&1.03& 1.07 &  $380 \pm 26$&1.33&0.99\\
	MgVII &	$278.40$ & $5.8$&$3240 \pm 90$&0.3& 0.35&  $65  \pm   10$&0.25&0.95\\       
	FeIX & $197.862$	& $5.9$&$380 \pm 12$&1.00& 1.13& $27 \pm    3$&0.97&1.09\\ 
	FeX	& $184.536$	& $6.0$&$1186  \pm    45$&0.89&0.81   &   $255   \pm   23$&0.94&0.58\\
	FeXI	& $180.401$	& $6.15$&$2270    \pm  140$&1.07&0.83  &  $ 1340   \pm   110$&1.11&0.76\\	
	FeXII	& $195.119$	& $6.20$&$1354   \pm   20 $&1.02&1.04  & $  1546   \pm   21$&1.09&1.06\\
	FeXIII	&$ 202.044$	& $6.25$& $1001  \pm    34 $&0.49& 0.57 & $  1650   \pm   40$&0.60&0.58\\
	FeXIV   &$ 211.318 $& $6.3$& $1120  \pm    100 $&0.85&1.15 & $   3420   \pm   170$&0.88&0.86\\	  
	FeXV	& $284.160	$&$ 6.35$&$3700  \pm    120$&1.13&1.08   &  $ 20070    \pm  280$&1.17&1.07\\
	FeXVI	& $262.984	$&$ 6.45$&$180   \pm   20   $&0.8&0.56& $  2870   \pm   80$&0.79&0.81\\	  
	SXIII	&$ 256.685	$& $6.45$&$399   \pm   44  $&0.49&0.37 & $  2380   \pm   100$&0.78&0.65\\	   
	CaXIV  & $193.866	$&$ 6.6$&$13   \pm   2  $&1.13&0.49 &$   512  \pm    13$&0.95&1.13\\      
	CaXV	& $200.972$ &$ 6.65$&$46  \pm    5  $&0.10& 0.05& $  650   \pm   20$&0.56&0.73\\	 
	CaXVII	& $192.858$ &$ 6.70$&$0  \pm    2  $&1.05& 0.99&$   540  \pm    10$&0.98&1.00\\     
\hline
\label{tab:eislines}
\end{tabular}
\end{center}
\end{table*}
\end{center}

\begin{figure}[t]
	\centering
	\includegraphics[scale= 0.6]{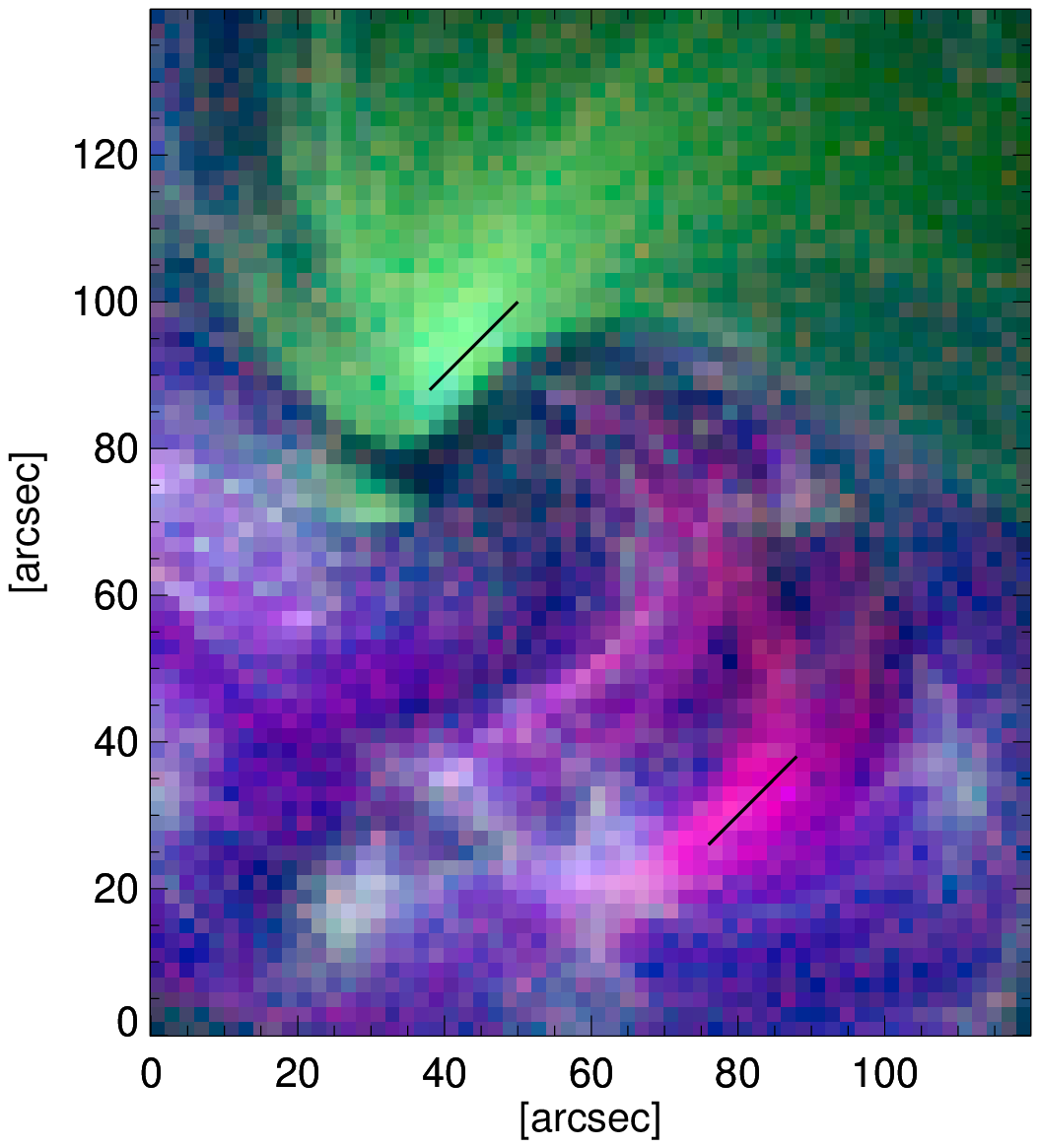}  
	\caption{Image of the active region in three AIA channels: 171 \AA\ (green), 335 \AA\ (blue) and  94 \AA\ (red). We analyse the two small regions marked by the strips, one is hot (pink), the other is cold (green), in more detail.}
	\label{fig:aia3col}
\end{figure}

Inside this field of view, we selected two small regions for more detailed analysis. In Paper I a special three-color coding is devised to highlight immediately hot and cool regions. In the three-color image (where each color is the intensity in a different AIA channel, green 171 \AA, blue 335 \AA, red 94 \AA), we selected a strip one-pixel wide and 7-pixels long, deep inside the hot region (where the hot part of the 94 \AA\ emission is high), and another (equal) one inside a colder region (high 171 \AA\ emission, Fig.\ref{fig:aia3col}).

\begin{figure*}[t]
	\centering
	\includegraphics[scale= 0.7]{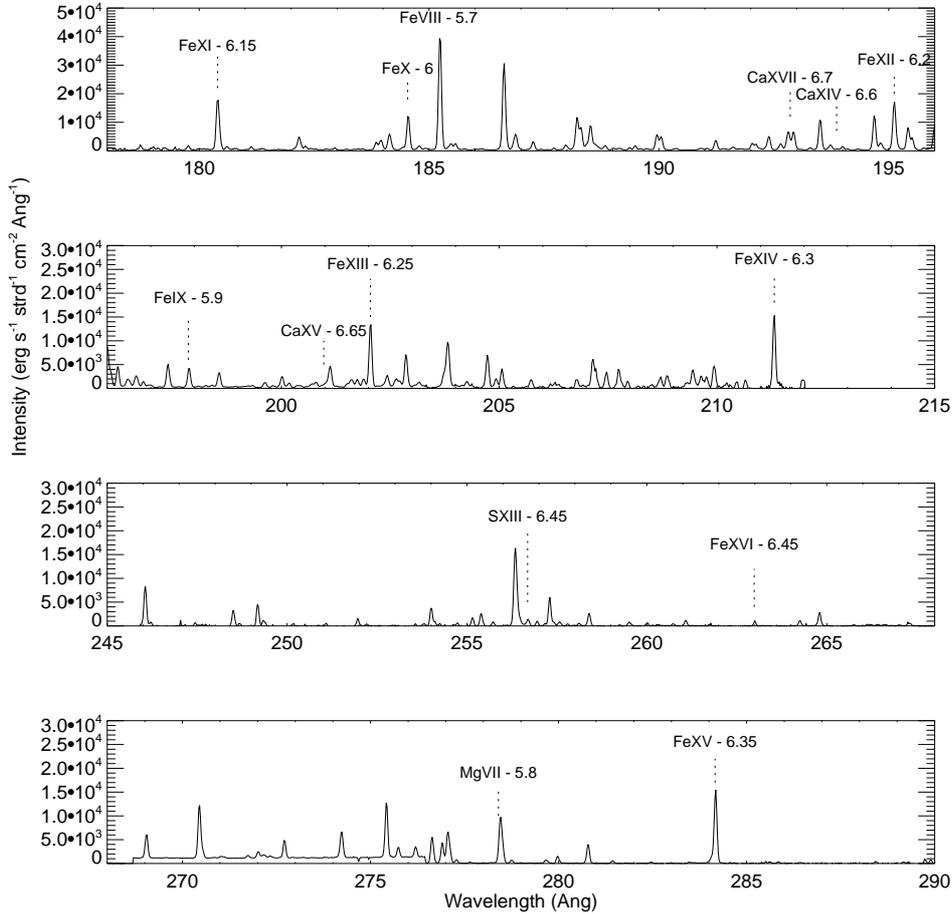}  
	\caption{Average Hinode/EIS spectrum over the strip in the cold region (green in Fig.~\ref{fig:aia3col}). The lines selected for further analysis are marked and labelled (with their temperature of maximum formation, $\log T$). }
	\label{fig:cold_spectrum}
\end{figure*}

\begin{figure*}[t]
	\centering
	\includegraphics[scale= 0.7]{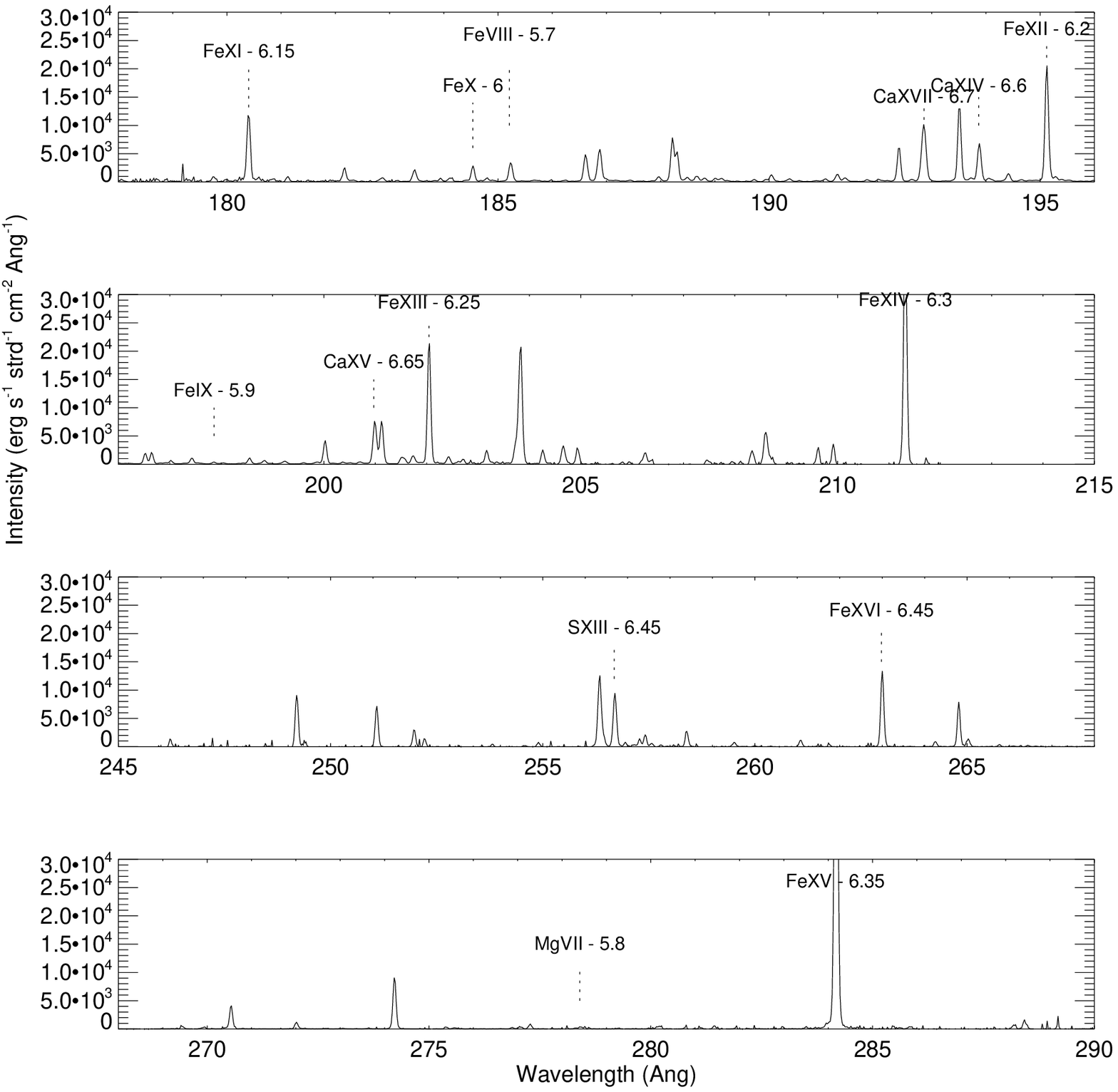}  
	\caption{As Fig.~\ref{fig:cold_spectrum} for the strip in the Hot Region (pink in Fig.\ref{fig:aia3col}).}
	\label{fig:hot_spectrum}
\end{figure*}

\begin{figure*}[t]
	\centering
	\includegraphics[scale= 0.7]{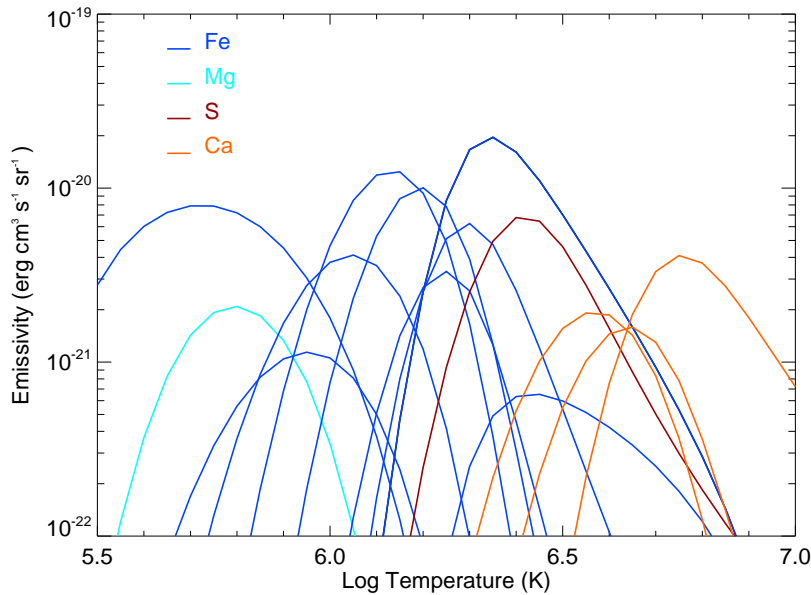}  
	\caption{Emissivity of the selected lines per unit emission measure.}
	\label{fig:eisemis}
\end{figure*}

We extracted EIS spectra in each pixel of the selected strips and, to increase the signal to noise ratio, we averaged the spectra over all the pixels in each strip. The resulting average spectra are shown in Fig.\ref{fig:cold_spectrum} and \ref{fig:hot_spectrum}. In these spectra the continuum is low and we see many emission lines for several elements, e.g. Fe, Ca, Mg.  We selected a subset of the emission lines with the following criterion. We conceptually divided the temperature range $5.5 < \log T < 7$ into bins $\Delta \log T \sim 0.1$ and choose approximately only one line that has the maximum formation temperature in a given bin, possibly the most intense one in at least one of the two spectra. We ended up with 14 lines, listed in Table~\ref{tab:eislines}, 9 from Fe ions, 3 from Ca, 1 from S and Mg, that provide a reasonably uniform coverage of the temperature range, as shown in Fig.~\ref{fig:eisemis}.  We use data from CHIANTI v 7.1 \citep{2013ApJ...763...86L}. 
We then fitted each line profile with a Gaussian (or multi-Gaussian for blended lines) and then we measured the flux by integrating the area below each Gaussian (Table~\ref{tab:eislines}). To each flux we applied the new EIS radiometric calibration, which also includes a 
correction for the long-term degradation of the EIS effective area \citep{2013A&A...555A..47D}. This leads to significantly higher (by a factor of about two) radiances of the EIS lines in the LW channel.
Our DEM modelling is mostly constrained by strong iron lines, for which the atomic data are reliable.
To a first approximation, we can therefore neglect any uncertainty due to atomic data and chemical abundances.
It is difficult to assess the accuracy of the new EIS calibration, so we decided to 
associate to each flux only the uncertainty due to photon statistics, and then use the results of the DEM modelling to discuss systematic errors (see Section~\ref{sec:discuss}). We measure zero flux for the Ca XVII line in the cold region, and the value in the table is the corresponding upper limit.

\subsection{DEM reconstruction}
\label{subsec:DEMrec}
 
 We use the radiances in the selected lines (Table~\ref{tab:eislines}) to derive the distribution of the emission measure vs temperature, DEM(T).  As a first step of the analysis we use the EM loci method \citep{1978PhDT.......298S}, by which each line intensity is divided by its emissivity. The loci of these curves represent, at each temperature, the maximum value of the  emission measure at that temperature, if all the plasma were isothermal.  Therefore, if the plasma were isothermal along the line of sight, all the curves would 
cross at a single point, exactly at the plasma temperature (see \citealt{2002A&A...385..968D} for more details). There are many inversion methods to obtain the DEM.

We applied the widely-used Markov-Chain Monte Carlo method (MCMC,  \citealt{1998ApJ...503..450K}). This technique is based on a Bayesian statistical formalism to determine the most probable DEM curves that reproduce the observed line intensities by Monte Carlo simulations. A very useful feature of this technique is the possibility to obtain an estimate of the uncertainty of the DEM in each temperature bin in which the DEM is computed (see e.g., \citealt{2011ApJ...728...30T}, \citealt{2012ApJ...758...54T} for more detail).

\begin{figure*}[h!]
	\centering
	\subfigure[Cold Region]{\includegraphics[scale= 0.6]{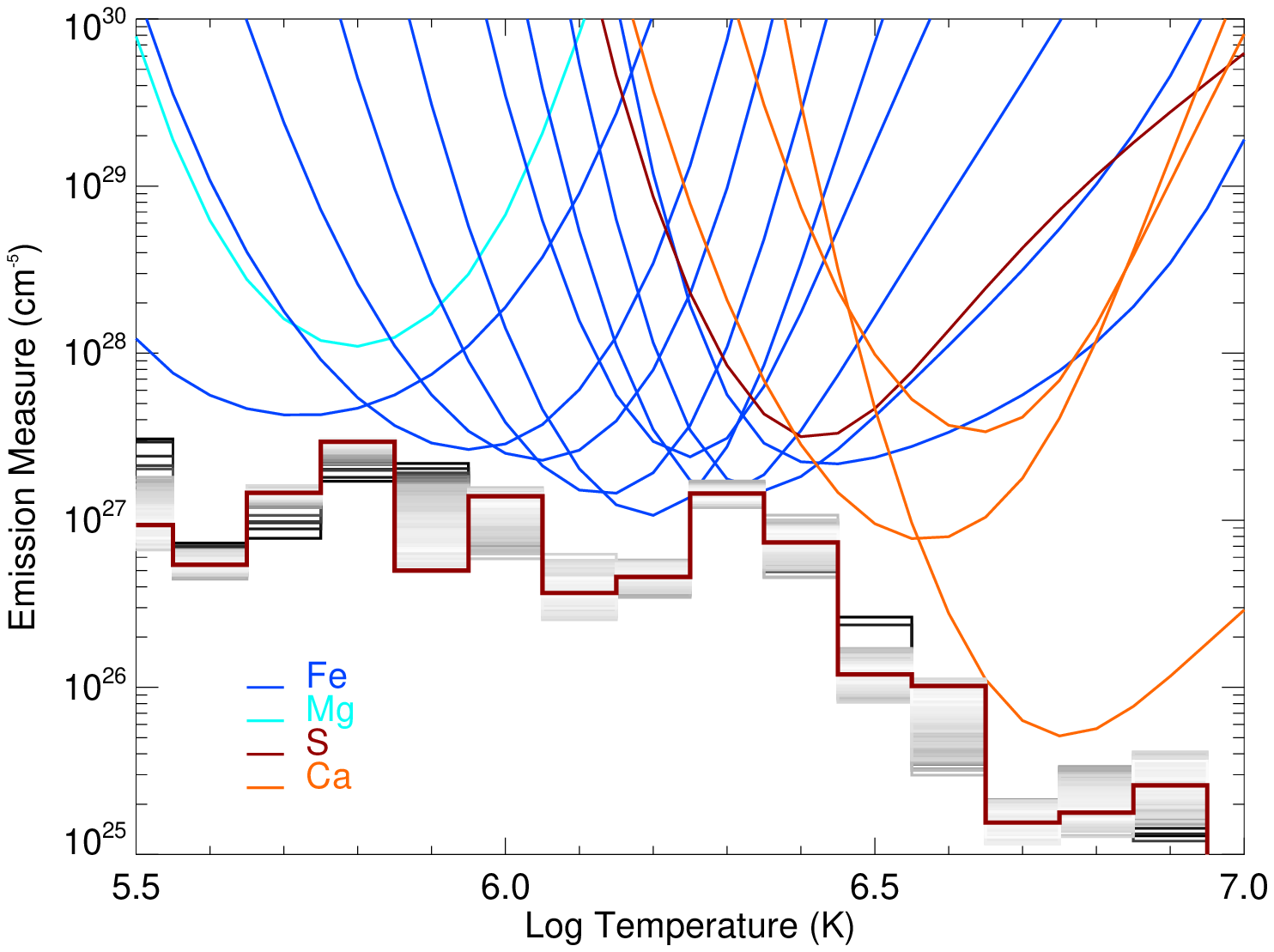}}
	\subfigure[Hot Region]{\includegraphics[scale= 0.6]{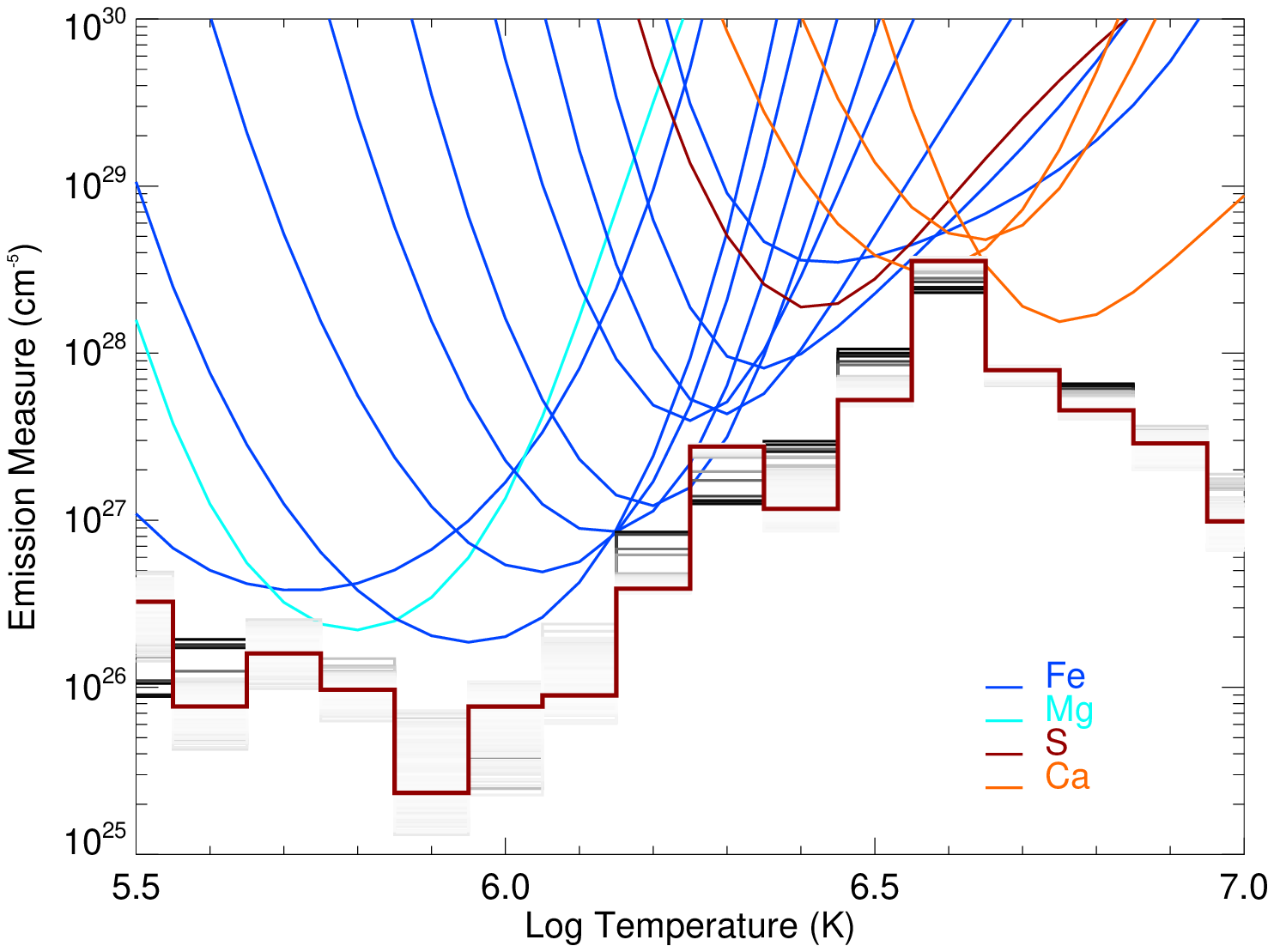}}
	\caption{Results of DEM reconstruction using the MCMC technique on 14 Hinode/EIS line fluxes (see text, and Table 1) for (a) the cold region, (b) the hot region, shown in Fig. \ref{fig:aia3col}. The red (histogram) curve is the solution that better reproduces the observed fluxes, the histogram cloud contains the solutions. The EM-loci curves are also shown for reference, each color marks a different element.}
	\label{fig:dem14}
\end{figure*}

For the application of the MCMC method, we assumed an electron density of $3 \times 10^{9}$ $cm^{-3}$, \cite{1992PhyS...46..202F} coronal abundances and CHIANTI 7.1  ionization equilibrium to compute the emissivity (per unit emission measure) of each line. The temperature range was divided into equal bins $\Delta \log T\sim 0.1$.  Fig.\ref{fig:dem14} shows the result of the reconstruction in which we added the EM-loci curves. We mark the most probable solution and the cloud of solutions in each temperature bin. The broader the cloud, the less constrained is the value of the best fit curve in that bin.

There is no small region in the plot where EM-loci curves all intersect; the plasma is therefore clearly multi-thermal both in the cold and in the hot segments. The information about the very hot plasma comes mostly from the Ca lines. The very low flux of the Ca XVII line in the cold region puts a strong constraint to reduce the emission measure for $\log T > 6.5$, while the flux is much higher in the hot region. The Fe and Ca lines show a good overall agreement while the Mg line leads to a higher emission measure in the cold region.

The MCMC method applied to the EIS data provides very different distributions in the cold and hot regions. In the cold region, the distribution is mostly flat for $\log T < 6.4$, then it decreases rapidly. The local peak at $\log T = 6.3$ might be significant, because of the small uncertainty. Overall the impression of a cool distribution is confirmed.  In the hot region, we see a different trend: the DEM increases monotonically for $\log T > 6$, with a slope between 2 and 3, it has a peak at $\log T = 6.6$ and then decreases gradually at higher temperature. This shape is not new for active region core \citep{2011ApJ...734...90W}, but our treatment of uncertainties makes the solutions better constrained. In particular, the components on the hot side of the DEM peak look quite well defined even with respect to the cooler side, where several lines have maximum formation temperatures. These hot emission measure components are at a level of $\sim 20$\% of the emission measure peak, therefore a significant fraction.

Table~\ref{tab:eislines} contains also the ratio of the line flux computed from the DEM reconstruction with the MCMC method to the observed one. Except for some well-known lines (Mg VII, Fe XIII) and for hot lines ($log T \geq 6.45$) in the cold region, not relevant for our analysis, good agreement (within 20\%) is found, comparable to that obtained in other similar analyses (e.g. \citealt{2011ApJ...734...90W}).

To try to better constrain the high-temperature part of the DEM,  first we add the information provided by the flux measured in the 94\AA\ AIA channel. As we briefly discussed previously, this band contains lines sensitive to emission from plasma at $\log T \sim 6.8$ (Fe XVIII),  but also lines sensitive to lower temperatures, so the intensity in the AIA band can be written as 

\begin{equation}
	I_{94} = EM_{cold}G_{94}(T_{cold})+EM_{hot}G_{94}(T_{hot})
	\label{eq:I94comps}
\end{equation}
where $EM_{cold}$ and $EM_{hot}$ are the emission measures of the cold and hot components, $G_{94}(T_{cold})$ and $G_{94}(T_{hot})$ are the values of the channel response functions of the cold and hot peaks respectively. 

We use the technique devised by \cite{2011ApJ...736L..16R} 
 to separate the two contributions and pick up the hot one. The technique uses the flux measured in the 171 \AA\ channel: 

\begin{equation}
	I_{171} = EM_{cold}G_{171}(T_{cold})
	\label{eq:I171}
\end{equation}

to constrain the cold component. Substituting Eq.\ref{eq:I171} in Eq.\ref{eq:I94comps} we obtain 

\begin{equation}
	I_{94} = \frac{I_{171}}{G_{171}(T_{cold})}G_{94}(T_{cold})+EM_{hot}G_{94}(T_{hot})
\end{equation}

We can then subtract the flux extrapolated from the 171 \AA\ channel from the 94 \AA\ flux. 
What is left is the hot part of the 94 \AA\ emission:

\begin{equation}
	I_{hot} = I_{94} - \frac{I_{171}}{G_{171}(T_{cold})}G_{94}(T_{cold})
\end{equation}

We note that a similar procedure, using the 171 and 193~\AA\ AIA bands was devised by 
\cite{2012ApJ...759..141W}. \cite{2013A&A...558A..73D}
suggested the use of the  171 and 211~\AA\ AIA bands instead, the latter one being 
used to estimate the contribution of the Fe XIV line to the 94~\AA\ band.
As shown in  \cite{2013A&A...558A..73D}, this method produces 
similar results as the method of \cite{2012ApJ...759..141W}.
The 94 \AA\ flux in the cold region is compatible with zero flux, so we put an upper limit as we did for the Ca XVII flux in the same region in Table~\ref{tab:eislines}.

After including the information from the AIA 94 \AA\ channel, the EM-loci AIA curves are in good agreement with the Ca XVII curves. In the hot region, the curves intersect both at $\log T \sim 6.6$ and at $\log T \sim 6.8$ but they are very similar in between. The DEM solutions derived with MCMC method are very similar to those obtained without the AIA flux, thus confirming coherent information.

\begin{figure*}[h]
	\centering
	\subfigure[Cold Region]{\includegraphics[scale= 0.6]{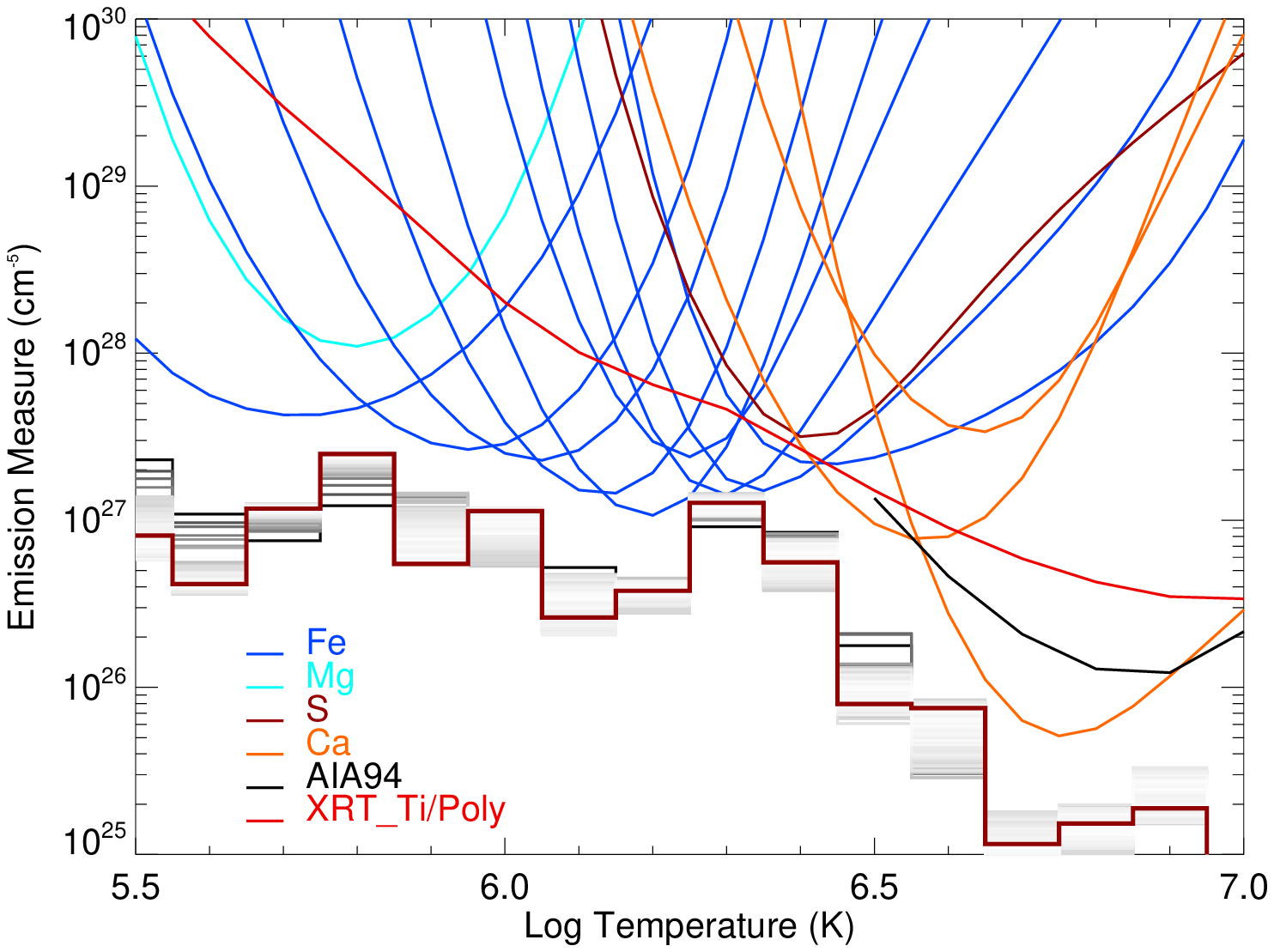}}
	\subfigure[Hot Region]{\includegraphics[scale= 0.6]{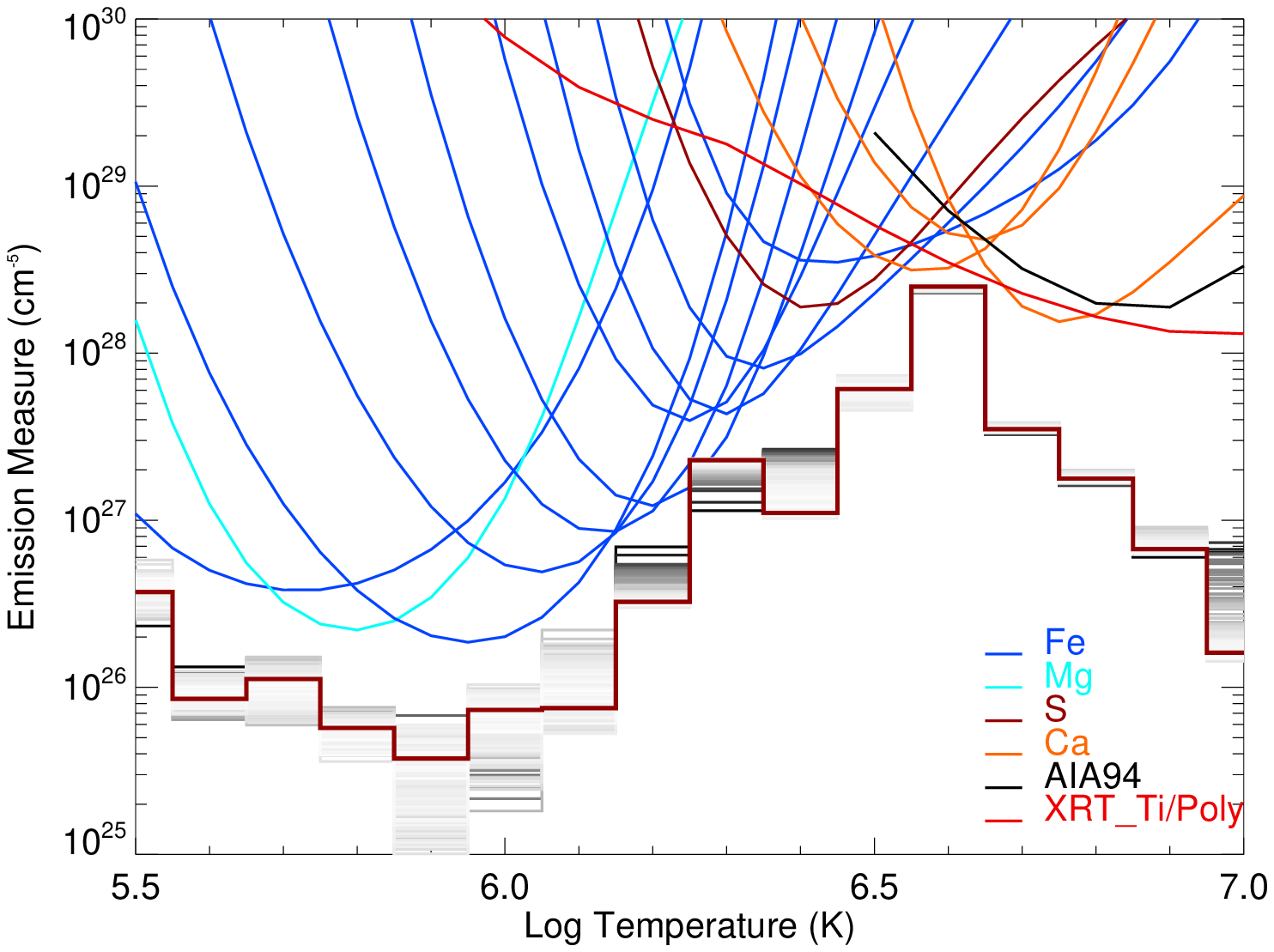}}
	\caption{Same as Fig.\ref{fig:dem14} adding the contribution of the fluxes measured in the "hot part" (see text) of the AIA 94 \AA\ channel (black line) and in the XRT Ty\_poly filter (red line).}
	\label{fig:dem14xrt}
\end{figure*}

Additional information about the hot components is independently available from the X-ray observation with Hinode/X-Ray Telescope (XRT). However, we expect looser constraints from the XRT filters because they are broadband, and have a broader temperature response.  We can simply plug in our analysis the flux measured in one XRT filter; in particular, we consider the Ti\_poly filter, which has the highest sensitivity to emission from plasma at $\log T \sim 6.9$. The result after including both AIA and XRT fluxes is shown in Fig.\ref{fig:dem14xrt}. We do not see qualitative difference from the DEM shown in Fig.~\ref{fig:dem14}, as expected from the broader AIA and XRT temperature responses (see EM-loci curve). A quantitative difference is that the DEM in the hot side of the DEM peak of the hot region is reduced by a factor $\sim 2$, i.e. to about 10\% of the peak, while maintaining a narrow cloud distribution for $\log T < 6.9$. 

\begin{center}
\begin{table*}[htb!]
\begin{center}
\caption{Fluxes measured in the AIA-94 \AA\ channel and in the XRT-Ti\_poly bands. 
For the 94\AA\ band we subtracted the cold component, as explained in subsection \ref{subsec:DEMrec}.}
\begin{tabular}{lcccc}
\hline\hline
Instrument&Cold region flux&Ratio&Hot region flux&Ratio\\
& (DN s$^{-1}$ pixel$^{-1}$)&MCMC & (DN s$^{-1}$ pixel$^{-1}$)&MCMC\\
\hline
AIA-94 \AA&$<0.4$&0.58 &$66 \pm 3$& 0.69 \\
XRT-TI\_poly&$53 \pm 5$& 0.94 &$2030	\pm 40$& 1.28\\ 
\hline
\label{tab:aiaxrtflx}
\end{tabular}
\end{center}
\end{table*}
\end{center}

\begin{figure*}[h!]
	\centering
	\subfigure[Cold Region]{\includegraphics[scale= 0.6]{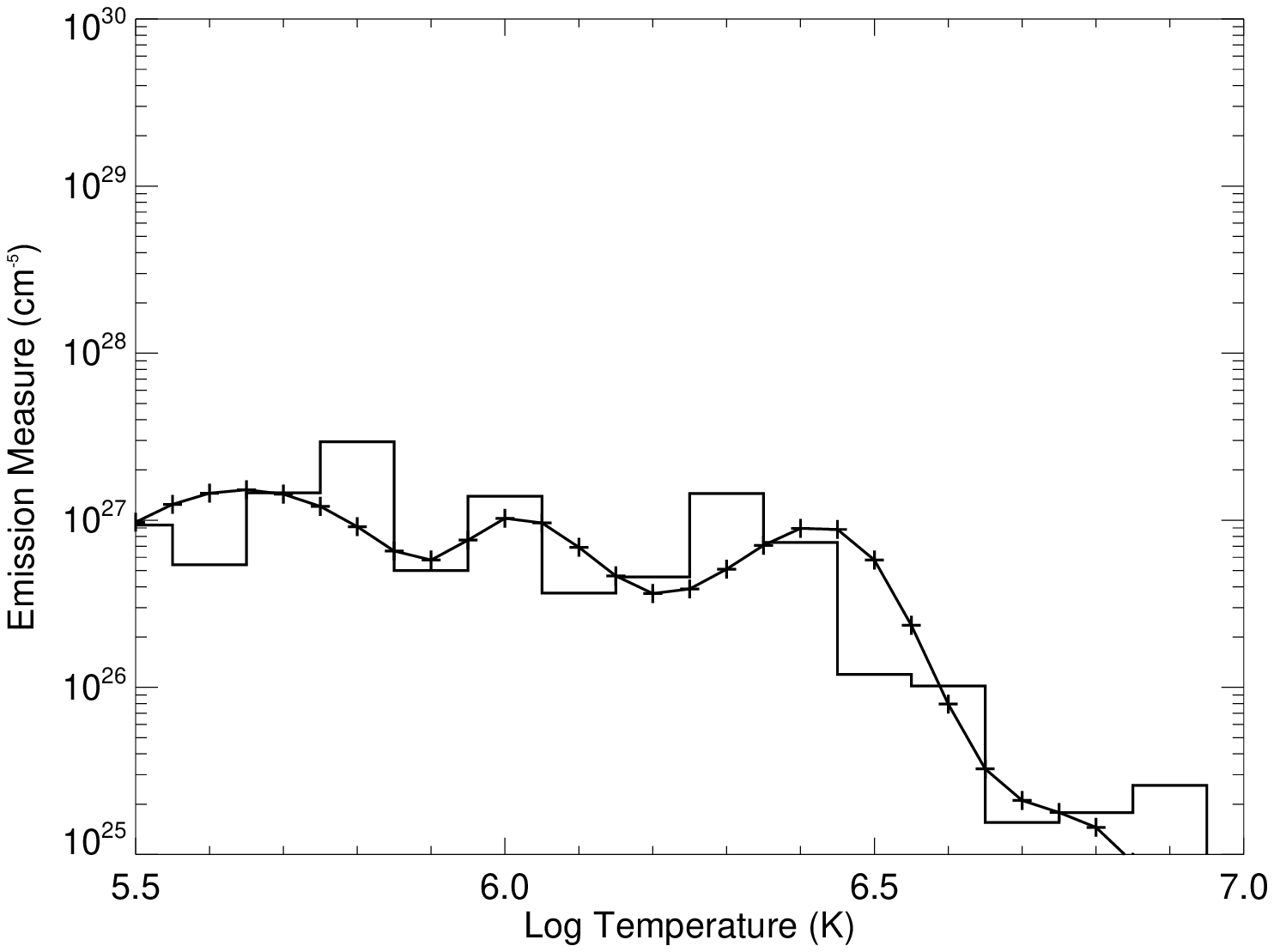}}
	\subfigure[Hot Region]{\includegraphics[scale= 0.6]{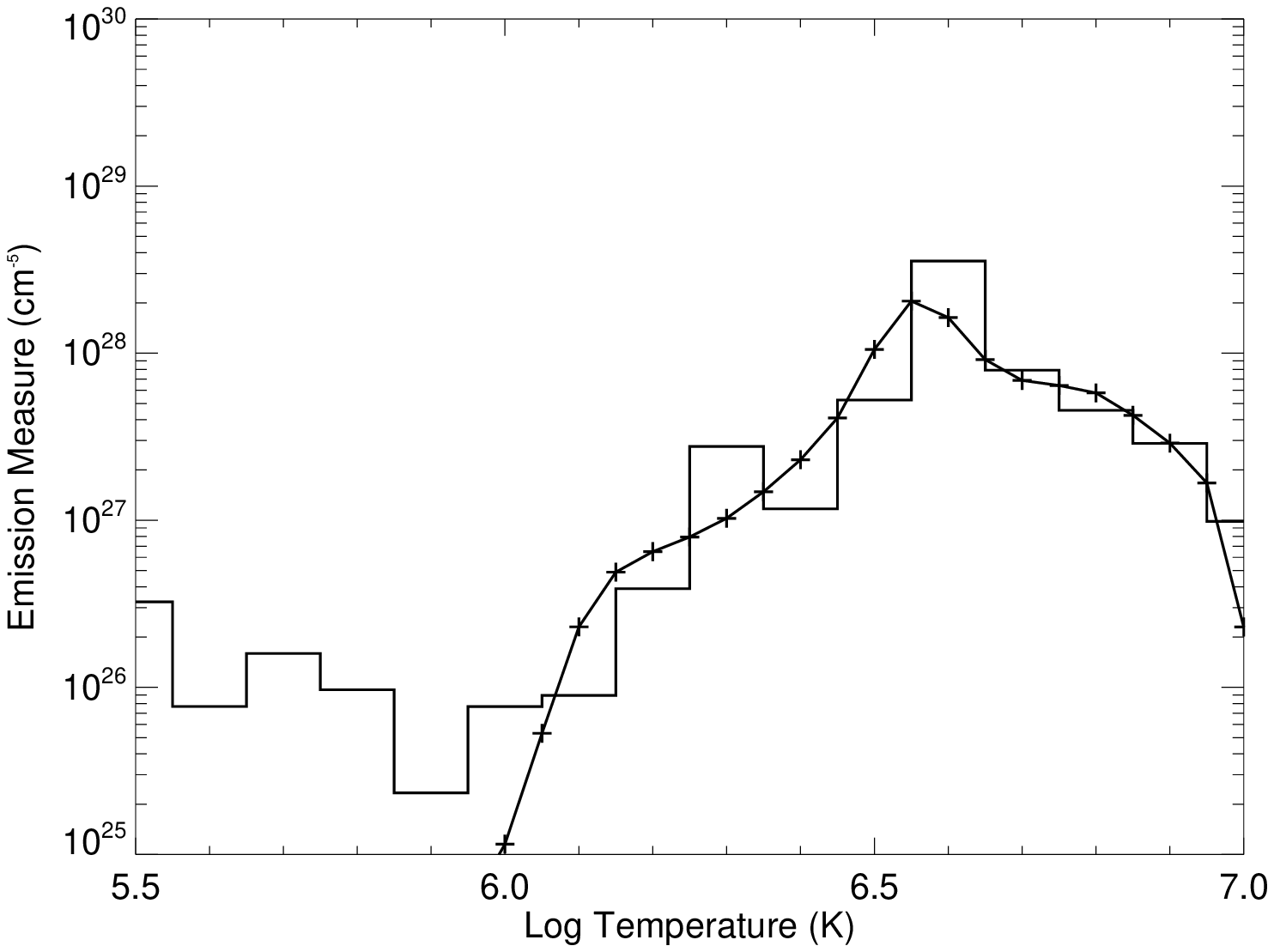}}
	\caption{Best DEM solutions from MCMC method (histogram, red in Fig.\ref{fig:dem14}) and from Del Zanna method (solid line $+$ symbols) for (a) the cold and (b) the hot region.  The Emission Measure values with MCMC are divided by the temperature to match the output from Del Zanna method.}
	\label{fig:petdzdem14}
	\end{figure*}

We further supported our analysis by comparing the best solutions of the DEM reconstruction on EIS data from the MCMC method with those from another method, devised by \cite{1999PhDT.........8D}. 
The method assumes that a smooth DEM distribution exists, and models it with a spline function.
The choice of the nodes of the spline is somewhat subjective, but the actual inversion 
is carried out following the maximum entropy method described in \cite{1991AdSpR..11..281M}.
We note that the same atomic data and input parameters were used for both inversions.  
The comparison is shown in Fig.~\ref{fig:petdzdem14}.

The results with the spline method are obviously smoother, but overall there is very good agreement between the two methods for both regions, especially in the temperature ranges of interest, and in particular in the hot part of the hot region. 
Good agreement (within 20\%) between observed and predicted intensities is found also with this method.

\section{Discussion and conclusions}
\label{sec:discuss}

We analysed the thermal distribution of the plasma along the line of sight in two different regions where the narrow-band imagers diagnosed very different dominant temperatures, a hotter and a cooler one. The analysis was mostly based on the spectral data from Hinode/EIS. Our major attention was to the hottest components that may be signature of impulsive heating at work, and here our aim was to reconstruct the whole EM distribution and check the coherence of the overall scenario, including the hot component. At variance from many other similar analyses, our approach was to consider a limited number of spectral lines ($\sim 1$ per temperature bin), and we picked the most intense lines uniformly covering the temperature range. In this way we simplified our control and interpretation of the reconstruction results, and all measured fluxes had the same weight in determining the global emission measure distribution. We also had a particular approach regarding the assessment of the uncertainties. The typical choice is to associate the same percentage error to all measured fluxes; 20\% is the most commonly assumed value (e.g. \citealt{2011ApJ...734...90W}). This uncertainty safely includes possible systematic unknown effects due to the instrument, atomic data and chemical abundances, but weights the same all measured values, independently of whether a line is strong or weak. Our choice is different. The measured fluxes were assigned exclusively the poissonian error, dictated only by the photon statistics. This allowed us to weight more the intense lines and to minimise the uncertainties of the solutions of the DEM reconstructions.

As a consequence, the DEM reconstruction with the MCMC technique leads to solution clouds with a narrow distribution around the best solution for many temperature bins (factor 2-3), significantly narrower than in previous works. Only a minority of temperature bins show a spread solution cloud. The constraint on the hot components is tighter. 

The error estimate is important in this analysis, and we are aware that the real uncertainties are surely larger than those that we assume. The uncertainties can influence considerably the error on the DEM solution, and sometimes even the solution itself, as thoroughly discussed in \cite{2012ApJ...745..111T}. We should also consider that the cloud spread may underestimate the real error on the DEM solution. The small uncertainties typically lead to a larger number of iterations with the MCMC method for better convergence \citep{2012ApJ...745..111T}, but our 400 iterations is certainly an appropriate number for our cases. Tests show that DEM structures are reliably recovered on scales larger than $\Delta \log T = 0.2$, so we do not discuss narrower features. 

The good agreement between predicted and observed line intensities confirms this, although we note that 
some significant discrepancies are present, in particular with the  Ca XV line, as also 
found previously \citep{2013A&A...558A..73D}. The good agreement between the two inversion methods,
already found in \cite{2011A&A...535A..46D}, is confirmed. This suggests that the main source of uncertainty resides in the 
choice of parameters, atomic data and instrument calibration.

Although our assumptions probably underestimate the errors, the coherent support from other instruments (AIA, XRT), the agreement with the results from another reconstruction method and the coherence with the morphology seen in the images spanning the different temperature regimes makes us confident that other errors should not affect our results considerably.

Overall, the analysis has confirmed the two general characteristics anticipated by the imagers. The comparison of the DEM distributions of the cold and hot regions has, on the one hand, revealed substantial thermal components for $logT < 6.3$ in the cold region, without showing prominent features. The reconstruction of the cold region also shows  minor components for $logT > 6.3$. Since the images show no bright features in the hot channels and lines, we may use the values of these emission measure components as sensitivity limits of our analysis, i.e. we may not trust emission measure components below $10^{26}$ cm$^{-5}$. 

The hot region shows a much more peaked thermal structure, with the positive gradient typically found in previous studies in coronal loops of active region cores \citep{2011ApJ...734...90W}. The peak is at a rather high temperature ($\log T = 6.6$) and beyond that the emission measure declines, but not very steeply and still showing a significant fraction of emission measure at temperature $\log T \leq 6.8$. Some components might be present at even higher temperature, although with higher uncertainties, up to the limit of the thermal sensitivity of our analysis ($\log T \sim 7$). We believe that the joint use of hot spectral lines, AIA 94 \AA\ channel and XRT filterbands, helps to partially remove the so-called "blind spot" for $\log T > 6.8$ \citep{2012ApJ...746L..17W}. Although the presence of the very hot component looks confirmed, still it is based on a very limited amount of information, and therefore some care should still be used. Some further feedback is provided by the comparison with the images, and between the images. The morphology of the region in the Ca XVII line only partially overlaps the morphology in immediately cooler lines (Fe XVI), while it matches well the X-ray and AIA 94 \AA\ morphology. 
This confirms that in the hot region selected in the present analysis there is indeed significant 
emission at 6~MK, resulting in strong  Ca XVII, so that the Fe XVIII emission in the  AIA 94 \AA\ band is also 
due to this component, unlike other cases in the cores of active regions where Ca XVII is not observed,
and the  Fe XVIII emission is due to a large emission measure at 3~MK \citep{2013A&A...558A..73D}.

Our analysis has remarked how spectral data are by far more constraining than the data from imagers, because the spectral lines are much more sensitive to temperature variations that the broader bands of the imagers (in agreement with the findings of \cite{2012ApJ...758...54T}. Still it stresses that a quantum leap in the diagnostics of the hottest DEM components needs constraints from more lines sensitive to emission from high temperature plasma. These might be, for instance, easily accessible to broad-band X-ray spectrometers, to which we look forward in future space missions.

% ***********************************************************

\acknowledgements
{
AP and FR acknowledge support from Italian Ministero dell’Universit\`a e Ricerca (MIUR). GDZ acknowledges support from STFC (UK).
PT was supported by contract SP02H1701R from Lockheed-Martin, NASA contract NNM07AB07C to SAO, and NASA grant NNX11AC20G. \emph{CHIANTI} is a collaborative project involving the \emph{NRL (USA)}, the \emph{Universities of Florence (Italy)} and
\emph{Cambridge (UK)}, and \emph{George Mason University (USA)}. Hinode is a Japanese mission developed and launched by ISAS/JAXA, with NAOJ, NASA, and STFC (UK) as partners, and operated by these agencies in cooperation with ESA and NSC (Norway). SDO data were supplied courtesy of the SDO/AIA consortium. SDO is the first mission to be launched for NASA’s Living With a Star Program. 
We thank the International Space Science Institute (ISSI) for hosting the International Team of S. Bradshaw and H. Mason: Coronal Heating – Using Observables to Settle the Question of Steady vs. Impulsive Heating.

}
%
% ***********************************************************

\bibliographystyle{aa}
\bibliography{corr}
\end{document}